\documentclass[twocolumn,superscriptaddress,showpacs,amsmath
,amssymb,aps,prl]{revtex4-1}
\usepackage{graphicx}
\usepackage{bm}
\usepackage{color}
\usepackage{hyperref}
\usepackage{verbatim}
\usepackage[caption=false]{subfig}
\usepackage{units}
\usepackage{braket}

\def\bh#1{black hole#1
  (BH#1)\gdef\bh{BH}}
\def\bbh#1{binary black hole#1
  (BBH#1)\gdef\bbh{BBH}}
\def\gw#1{gravitational wave#1
  (GW#1)\gdef\gw{GW}}
\def\nr#1{numerical relativity#1
  (NR#1)\gdef\nr{NR}}
\def\pnw#1{post-Newtonian#1
  (PN#1)\gdef\pnw{PN}}
\def\GT#1{Georgia Tech#1
  (GT#1)\gdef\GT{GT}}

\begin{document}

\title{Coping with Junk Radiation in Binary Black Hole Simulations}

\author{Kenny Higginbotham}
\affiliation{Center for Relativistic Astrophysics,
School of Physics,
Georgia Institute of Technology, Atlanta, GA 30332}

\author{Bhavesh Khamesra}
\affiliation{Center for Relativistic Astrophysics,
School of Physics,
Georgia Institute of Technology, Atlanta, GA 30332}

\author{Jame P. McInerney}
\affiliation{Center for Relativistic Astrophysics,
School of Physics,
Georgia Institute of Technology, Atlanta, GA 30332}

\author{Karan Jani}
\affiliation{Center for Relativistic Astrophysics,
School of Physics,
Georgia Institute of Technology, Atlanta, GA 30332}

\author{Deirdre M. Shoemaker}
\affiliation{Center for Relativistic Astrophysics,
School of Physics,
Georgia Institute of Technology, Atlanta, GA 30332}

\author{Pablo Laguna}
\affiliation{Center for Relativistic Astrophysics,
School of Physics,
Georgia Institute of Technology, Atlanta, GA 30332}

\begin{abstract} 
Spurious junk radiation in the initial data for binary black hole numerical simulations has been an issue of concern. The radiation affects the masses and spins of the black holes, modifying their orbital dynamics and thus potentially compromising the accuracy of templates used in gravitational wave analysis. Our study finds that junk radiation effects are localized to the vicinity of the black holes. Using insights from single black hole simulations, we obtain fitting formulas to estimate the changes from junk radiation on the mass and spin magnitude of the black holes in binary systems. We demonstrate how these fitting formulas could be used to adjust the initial masses and spin magnitudes of the black holes, so the resulting binary has the desired parameters after the junk radiation has left the computational domain. A comparison of waveforms from raw simulations with those from simulations that have been adjusted for junk radiation demonstrate that junk radiation could have an appreciable effect on the templates for LIGO sources with SNRs above 30. 
\end{abstract}


\maketitle

\emph{Introduction:} This letter presents a method to deal with the spurious \emph{junk} radiation present in puncture-type initial data for \bbh{} simulations.

When Einstein's theory of General Relativity is viewed as an initial-value problem, the initial data consist of the spatial metric $\gamma_{ij}$ and the extrinsic curvature $K_{ij}$ of the initial space-like hypersurface in the space-time foliation. In the pair $\lbrace \gamma_{ij}, K_{ij}\rbrace$, not everything is freely specifiable. Four \emph{pieces} are fixed by the Hamiltonian and momentum constraints.  The York-Lichnerowicz approach~\cite{2010nure.book.....B,1979sgrr.work.....S} provides a path to identify those four pieces via conformal transformations and tranverse-traceless decompositions. With this approach, after assuming conformal flatness ($\gamma_{ij} = \Phi^4\eta_{ij}$) and vanishing trace of the extrinsic curvature ($K^i_i = 0$), the Hamiltonian constraint for vacuum space-times reads: 
\begin{equation}
\tilde\Delta \Phi
+ \frac{1}{8}\Phi^{-7} \tilde A_{ij} \tilde A^{ij} = 0 \label{eq:HamCF}\,.
\end{equation}
Here, tildes denote tensors and operators in conformal space, and $\tilde A_{ij}$ is the conformal trace-free extrinsic curvature satisfying $\widetilde\nabla_i \tilde A^{ij} = 0$, namely the momentum constraint.

To construct puncture-type initial data representing \bbh{s}, one uses the Bowen-York ~\cite{1980PhRvD..21.2047B} point-source solutions to $\widetilde\nabla_i \tilde A^{ij} = 0$: \begin{eqnarray}
  \tilde A^{ij} &=& \frac{3}{2\,r^2}\left[ P^i l^j + P^jl^i - ( \eta^{ij}-l^il^j) (P^k l_k)\right]\label{eq:KP}\\
  \tilde A^{ij} &=& \frac{6}{r^3}l^{(i}\epsilon^{j)kl}S_kl_l  \label{eq:KS}
\end{eqnarray}
with $l^i = x^i/r$ and $\epsilon^{ijk}$ the Levi-Civita symbol. Also, $P^i$ and $S^i$ are the linear and angular momentum of the point-source, respectively. Given (\ref{eq:KP}) and (\ref{eq:KS}), Eq.~(\ref{eq:HamCF}) is solved using the \emph{puncture} approach introduced by Brandt and Br\"ugmann~\cite{Brandt:1997tf}. The essence of this approach is to factor out the \bh{} singularity, namely  
\begin{equation}
\Phi = 1 + \frac{m_1}{|r - r_1|} + \frac{m_2}{|r - r_2|} + u
\end{equation}
with $r_{1,2}$ the locations of the \bh{s} and $u$ a regular function. The parameters $m_{1,2}$ are commonly referred as the \emph{bare} or puncture masses. Ansorg~\cite{Ansorg:2004ds} developed an elegant solver for Eq.~(\ref{eq:HamCF}) based on spectral methods called \texttt{2Punctures}. The solver is one of the most widely used codes by the \nr{} community.

To construct astrophysically relevant \bbh{} initial data, one thus needs to provide the values for the parameters $m_{1,2}$, $r_{1,2}$, $P^i_{1,2}$ and $S^i_{1,2}$. The most common approach for this is with the assistance of \pnw{} approximations. \pnw{} equations of motion are used to evolve the \bbh{} of interest from large separations until the separation at which the \nr{} evolution will start. The parameters of the \bbh{} at the end of the \pnw{} evolution are used as input parameters to solve Eq.~(\ref{eq:HamCF}). There is a subtlety here. The bare puncture masses $m_{1,2}$ are not the masses of the \bh{s}. As first guess for these parameters, one uses the \pnw{} \bh{} masses, but iterations are needed to adjust the puncture bare masses until the masses of the \bh{s} match the desired \pnw{} masses.

There is an issue with the puncture  \bbh{} initial data as just described.  For sufficiently large binary separations, one would expect the space-time in the neighborhood  of each \bh{} to be close to a boosted Kerr solution, or boosted Schwarzschild solution if the \bh{} is not spinning. This is not the case for the puncture data with Bowen-York  extrinsic curvatures, not even for a single \bh{.} The reason for this are the conformal flatness assumption and the Bowen-York  extrinsic curvatures. The space-time of a boosted or a Kerr \bh{} is not conformally flat. Also, the Bowen-York  extrinsic curvatures are not the extrinsic curvatures for a boosted or Kerr \bh{.} As a consequence, the puncture Bowen-York initial data contain spurious or \emph{junk} radiation. 
  
\begin{figure}
    \centering
    \includegraphics[width=0.4\textwidth]{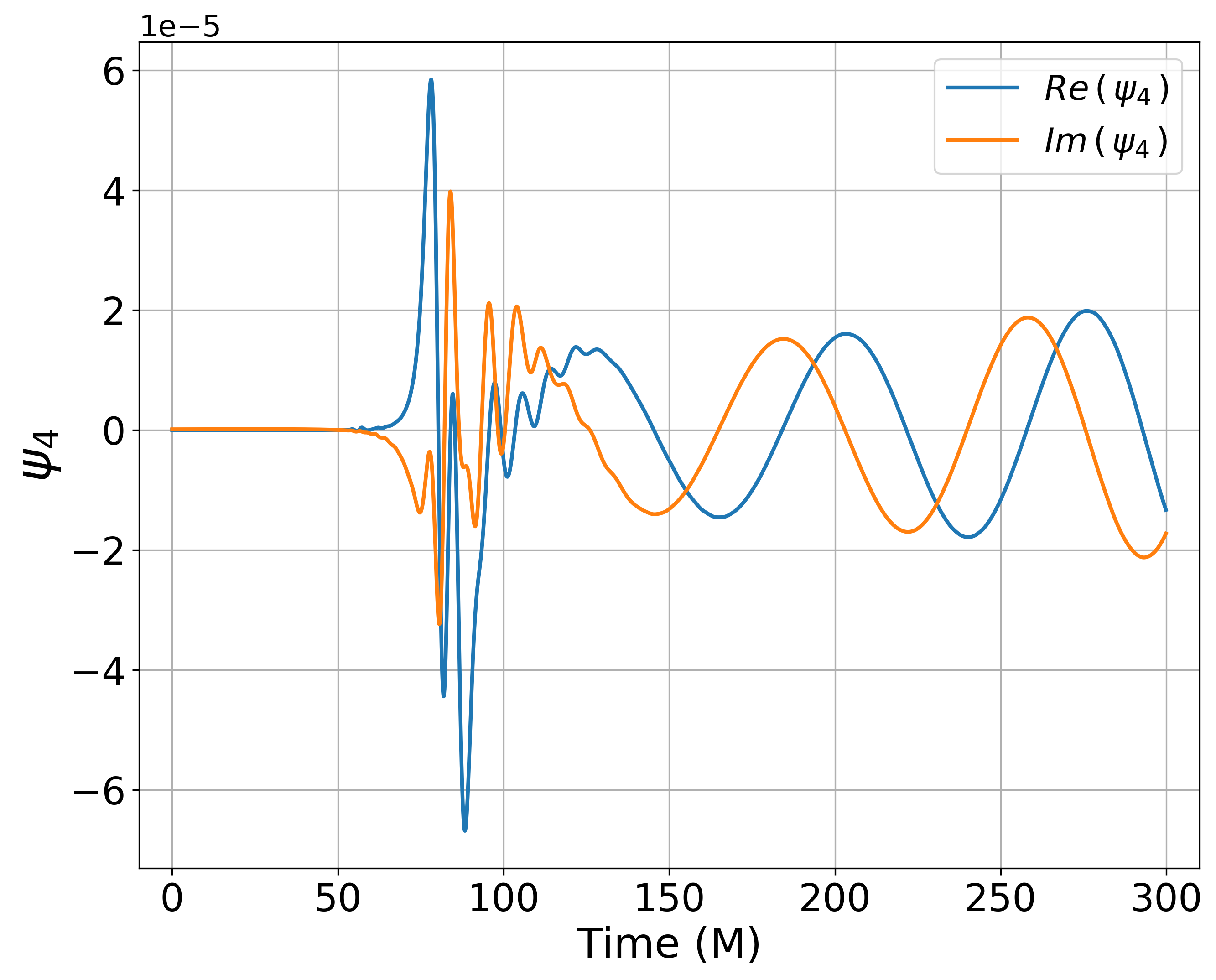}
	\caption{The real and imaginary parts of the $l=2$, $m=2$ modes of Weyl scalar $\Psi_4$ extracted at a radius $75\,M$ for the wavefrom GT0860 in the Georgia Tech catalogue. This is an equal mass, precessing BBH system with spin magnitude of a=0.8. The junk radiation is evident at the begining and appears to end around a time $134\,M$, with $M$ the total mass of the binary.}
    \label{fig:psi4}
\end{figure}

In numerical evolutions, junk radiation manifests itself as a burst.
Figure~\ref{fig:psi4} depicts an example in terms of the Weyl scalar $\Psi_4$ as a function of time for the
$l=2$, $m=2$ mode. There is an ongoing debate in the NR community about the extent to which the junk radiation introduces appreciable  changes to the binary, specifically changes to the spins and masses of the \bh{s}, and thus to the orbital dynamics of the binary~\cite{2007CQGra..24S..15H,2009CQGra..26k4002L,2009PhRvD..80l4039J,2009PhRvD..80l4051C,2010CQGra..27k4005K,2013PhRvD..88h4033Z,2018PhRvD..98d4014S,2015PhDT........97R,2011PhRvD..84h4038S}. Some of the studies attempt to tame the junk radiation by moving away from conformal flatness~\cite{2009CQGra..26k4002L,2009PhRvD..80l4039J}, others introduce explicitly \pnw{} corrections~\cite{2010CQGra..27k4005K}.
Our view here is to obtain first a detailed characterization of the effects from junk radiation on the holes and then introduce adjustments in the input parameters of the binary that anticipates the changes produced by the junk radiation. The expectation is that,  after the junk radiation dissipates away, one is left with the \bbh{} system one originally intends  to have.  

\begin{figure}
    \centering
	 \includegraphics[trim = {0 0 0 -10},width=0.42\textwidth]{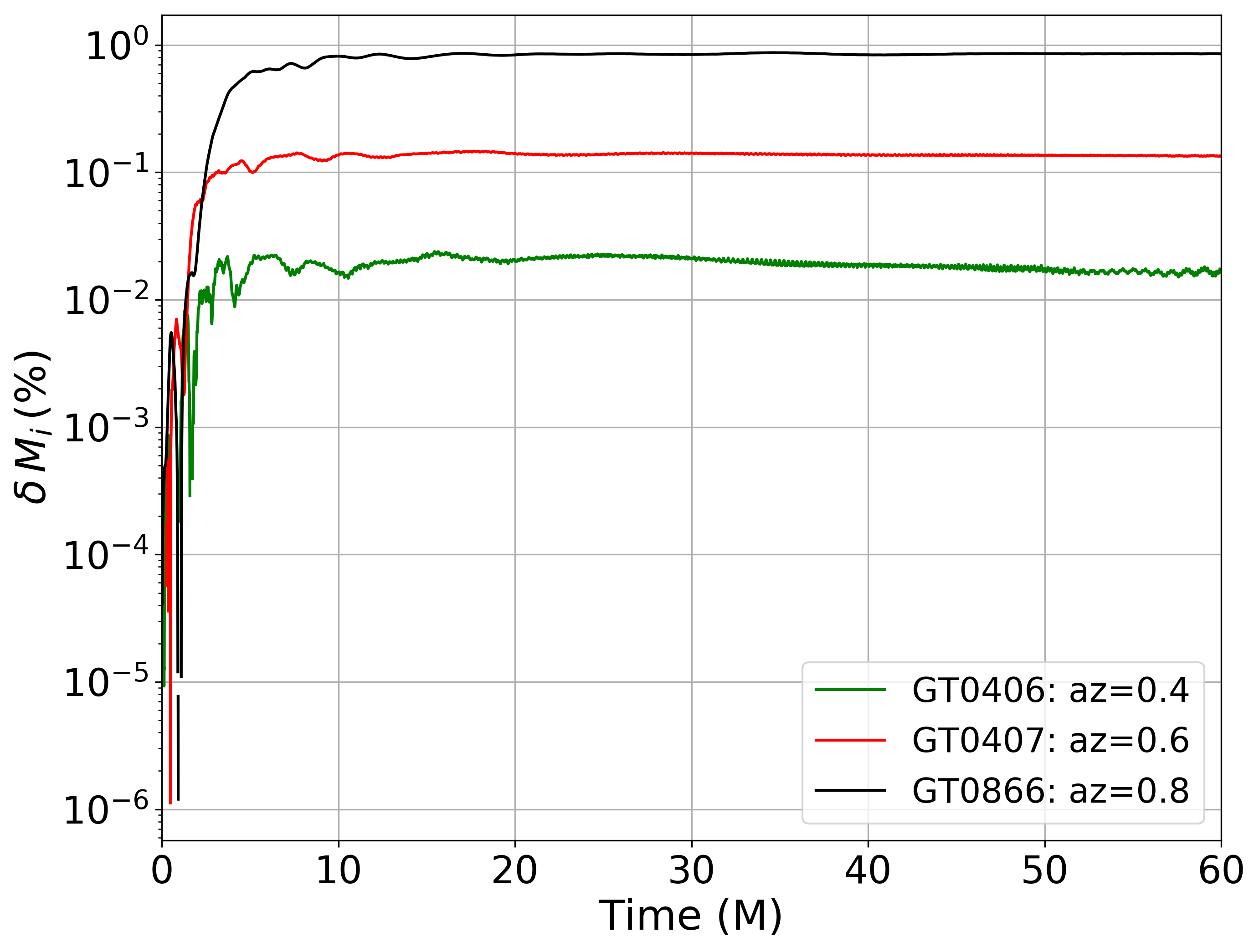}
    \caption{The percent change in initial irreducible mass of the \bh{s} for three simlations from the Georgia Tech catalogue: GT0406, GT0407 and GT0866. An increase in the mass is seen over a period of approximately $20M$.}
    \label{fig:mass_change}
\end{figure}

\emph{Waveform Analysis:} Our work is based on the Georgia Tech catalogue of \bbh{} simulations~\cite{Jani:2016wkt}. The first step we took was to monitor in our simulations the behavior of the masses and spins of the \bh{s} in a window between the initial time of the simulation and the end of the burst of junk radiation.  Figure~\ref{fig:mass_change} shows the percent change $\delta M_i$ in the initial irreducible mass of the \bh{s} as a function of time for three binary simulations from the catalogue: GT0406, GT0407 and GT0866. The irreducible mass $M_{i}$ is computed from the area $A$ of \bh{} horizon: $M_{i}  = \sqrt{A/4\,\pi}$. It is evident in Fig.~\ref{fig:mass_change} the increase in the masses of the \bh{s} during the first $20\,M$ of the simulation. This trend was observed in all the simulations for which we tracked the masses of the holes. Similar jumps were also observed in the spins.

\emph{Learning from single black hole simulations:} Next we investigated whether the jumps in mass and spin were due to \emph{local} effects in the neighborhood of the \bh{s} or if they involved correlations between the holes in the binaries. To answer this question, we looked at the effects of junk radiation on a single \bh{.} We carried out simulations expanding the dimensionless spin parameter $a = S/M_h^2$ in the range $0 \le a \le 0.8$ and the speed $v= P/M_h$ in the range $0 \le v \le 0.3$, with $S$ the angular momentum, $P$ linear momentum, and $M_h$ the mass of the \bh{}, where $M_h^2 = M_i^2+S^2/(2M_i)^2$. The spin of the \bh{} was aligned with the $z$-direction and the momentum with the $y$-direction. To a good approximation, these configurations cover the initial setups of \bh{s} in the non-precessing \bbh{} simulations in our catalogue. 

Not surprisingly, the single \bh{} simulations also showed increases in mass and spin. Furthermore,  those increases  took place, as with the \bbh{} simulations, during  the first $20M$ of the simulation time. This suggests that the effects of junk radiation are localized near the hole and do not depend on the presence of the other hole.

\begin{figure*}
    \centering
\subfloat[]  {\includegraphics[width=0.5\textwidth]{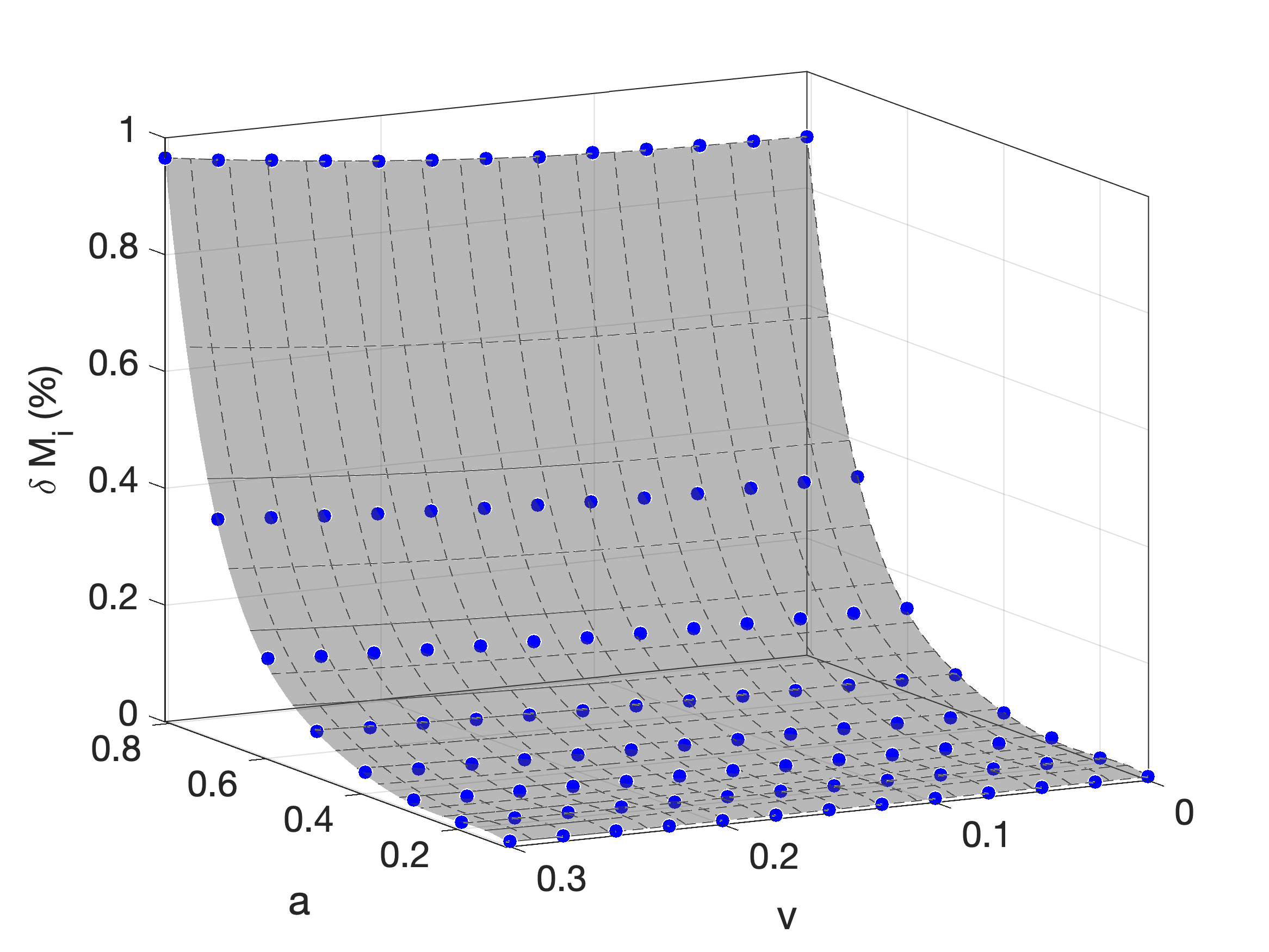}}
\subfloat[]  {\includegraphics[width=0.5\textwidth]{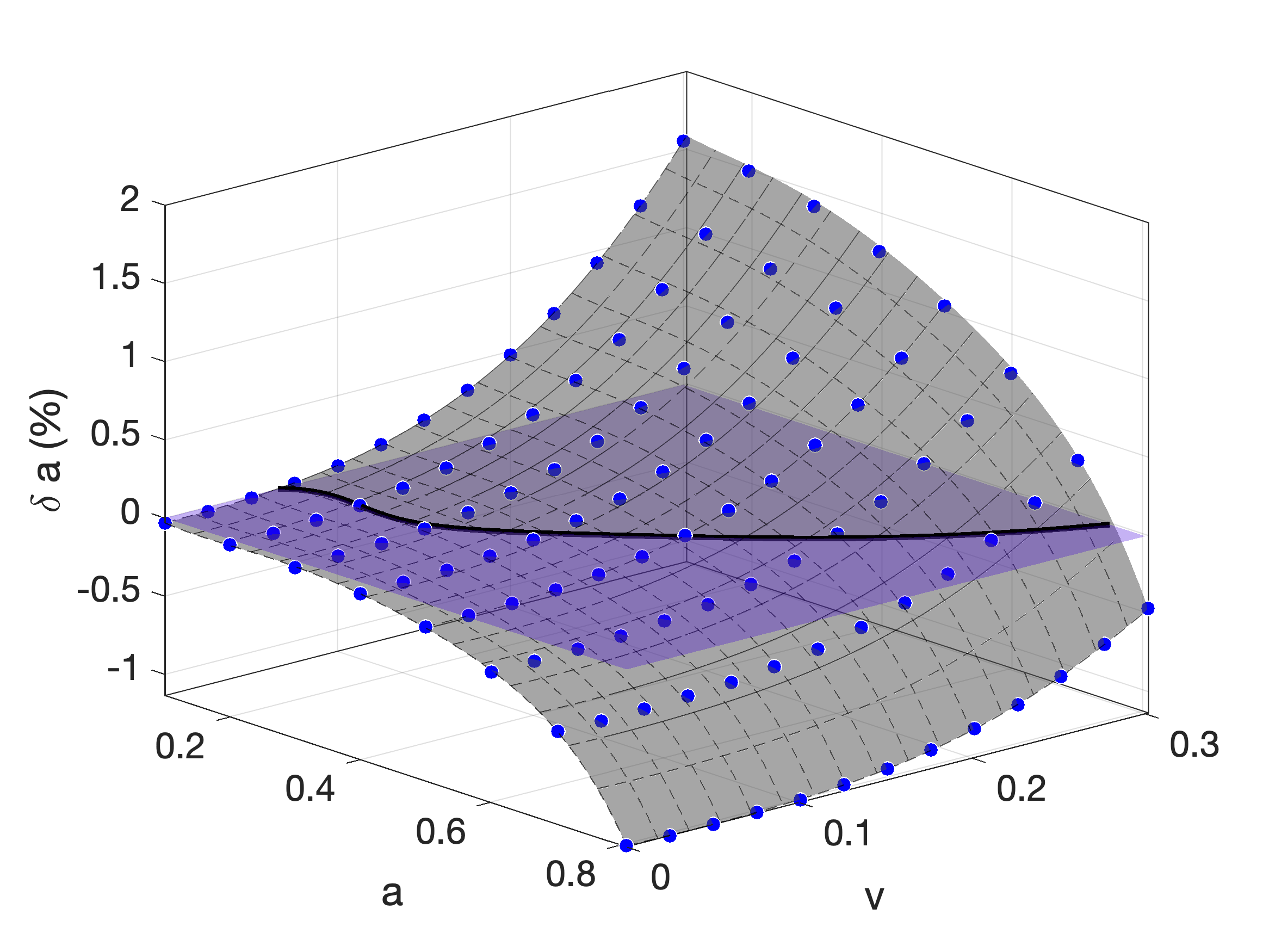}}
    \caption{Percent change in the (a) irreducible mass of the \bh{} $M_i$  and (b) dimensionless spin $a$  in the single \bh{} simulations as a function of $a$ and $v$ (blue dots). Two-parameter fits to the data by Eq.~(\ref{eq:fit})  are shown as a grey surfaces.}
    \label{fig:fit_mass_SBH}
\end{figure*}

Figure~\ref{fig:fit_mass_SBH} shows with points the percentage change in the \bh{} irreducible mass $\delta M_i$ and  $\delta a$ for all the single \bh{} simulations as a function of the dimensionless spin $a$ and speed $v$. The grey surfaces in Figure~\ref{fig:fit_mass_SBH} are fits to the data where we use the following fitting function: 
\begin{eqnarray}
 F  
&=& c_{00} + c_{10}a + c_{20}a^2 +c_{30}a^3+ c_{40}a^4+ c_{50}a^5 \nonumber \\
&+& c_{01}v + c_{02}v^2 + c_{03}v^3+ c_{04}v^4+ c_{05}v^5 \nonumber\\
&+& c_{11}av + c_{12}av^2 +c_{13}av^3+ c_{14}av^4\nonumber \\
&+& c_{21}a^2v +c_{31}a^3v+ c_{41}a^4v\nonumber \\
&+& c_{22}a^2v^2 + c_{23}a^2v^3 + c_{32}a^3v^2 
    \label{eq:fit}
\end{eqnarray}
The coefficients for the fit to $\delta M_i$ and $\delta a$ are given in Table~\ref{table:fit}.

Interestingly, from Fig.~\ref{fig:fit_mass_SBH}a, the effect of junk radiation on the mass correlates stronger with the initial spin than with the speed of the puncture. This is not the case with the effect on the spin of the \bh{.} As it can be seen from Fig.~\ref{fig:fit_mass_SBH}b, the junk radiation reduces the spin for larger initial spin and increases the spin for larger speeds. As a consequence, there is a family of cases for which the effects cancel out. These are the cases when the surface in Fig.~\ref{fig:fit_mass_SBH}b intersects the $\delta a = 0$ plane, and they are denoted with the black line. Another important finding was that the junk radiation only affected the magnitude of the spin but not its direction. 

\begin{table}[]
\begin{tabular}{|c|c|c|c|c|c|}
\hline
$c_{ij}$ & $\delta M_i$ & $\delta a$ &$c_{ij}$ & $\delta M_i$ & $\delta a$ \\
\hline\hline
$c_{00}$  &   -0.09935  & 0.1294	   &$c_{40}$  &  -65.83 &79.59   \\
$c_{10}$  &   2.148  & -3.169   &$c_{31}$  &   -4.27  & 13.47 \\ 
$c_{01}$  &    0.05612& -0.0977& $c_{22}$  &  -1.565  &  -5.56 \\
$c_{20}$  &     -14.92  &   21.01& $c_{13}$  &    3.918  & -13.17\\
$c_{11}$  &     -0.4674  & 1.118& $c_{04}$  &   -21.74  &  288.9  \\
$c_{02}$  &  -0.6845 &   7.696 &$c_{50}$  &     36.64   &  -40.72\\
 $c_{30}$  &           46.3   &  -61.24  &$c_{41}$  &   2.524 &     -7.763\\
 $c_{21}$  &     2.386  &  -7.675& $c_{32}$  &  2.931   & 2.056 \\
$c_{12}$  &      -0.2141  & 2.781& $c_{23}$  &   -0.7278  &      -35.24 \\
$c_{03}$  &   5.791 & -15.01&$c_{14}$  &  -5.623    &  12.7  \\
&&& $c_{05}$  & 29.78 & -413.8  \\
\hline
\end{tabular}
 \caption{Fitting coefficients for $\delta M_i$ and $\delta a$ in Eq.~\ref{eq:fit}.}
\label{table:fit}
\end{table}
   
\begin{figure*}
    \centering
\subfloat[]{    \includegraphics[width=0.47\textwidth]{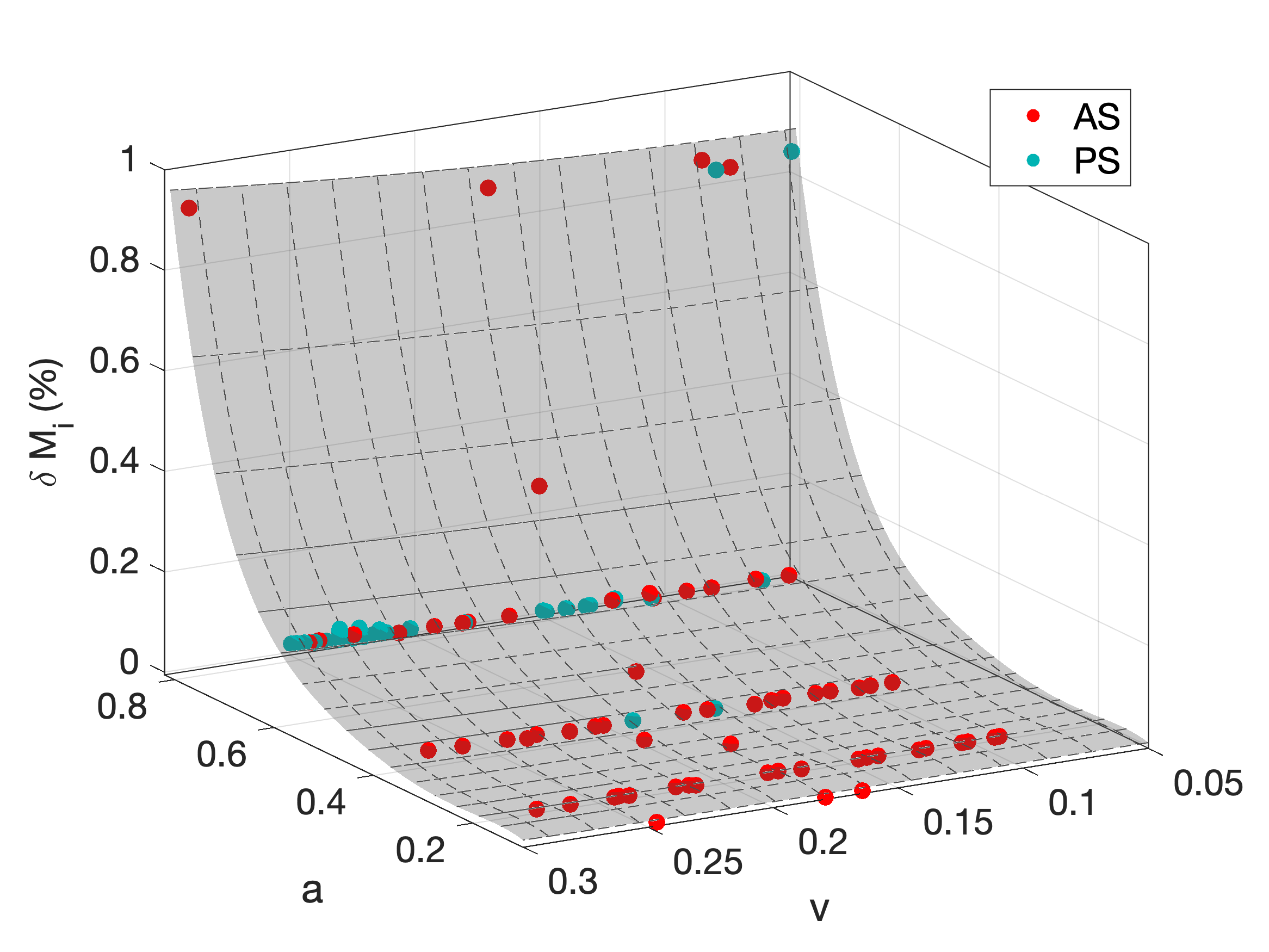} }
\subfloat[]{       \includegraphics[width=0.47\textwidth]{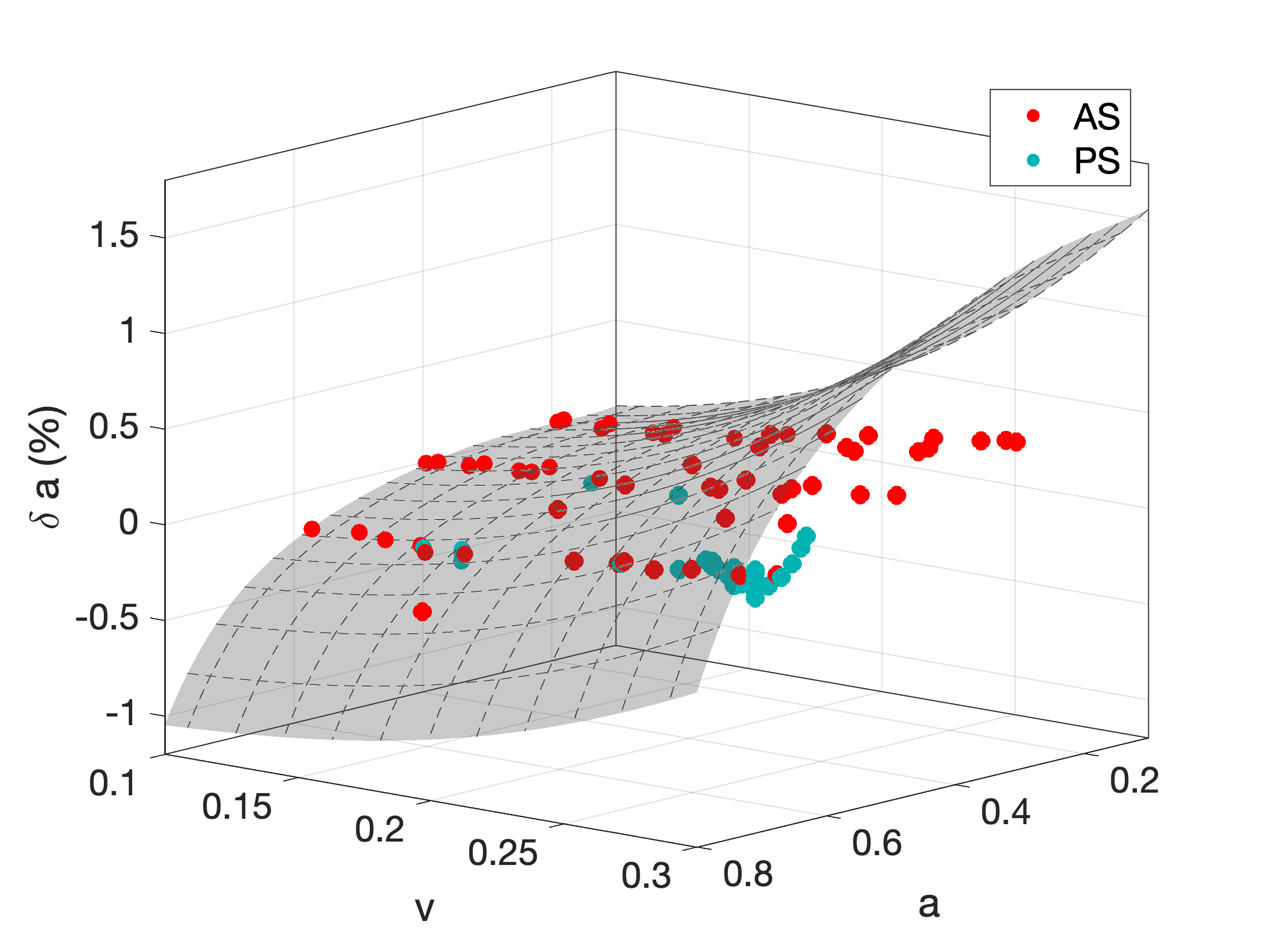}}\\
\subfloat[]{       \includegraphics[width=0.47\textwidth]{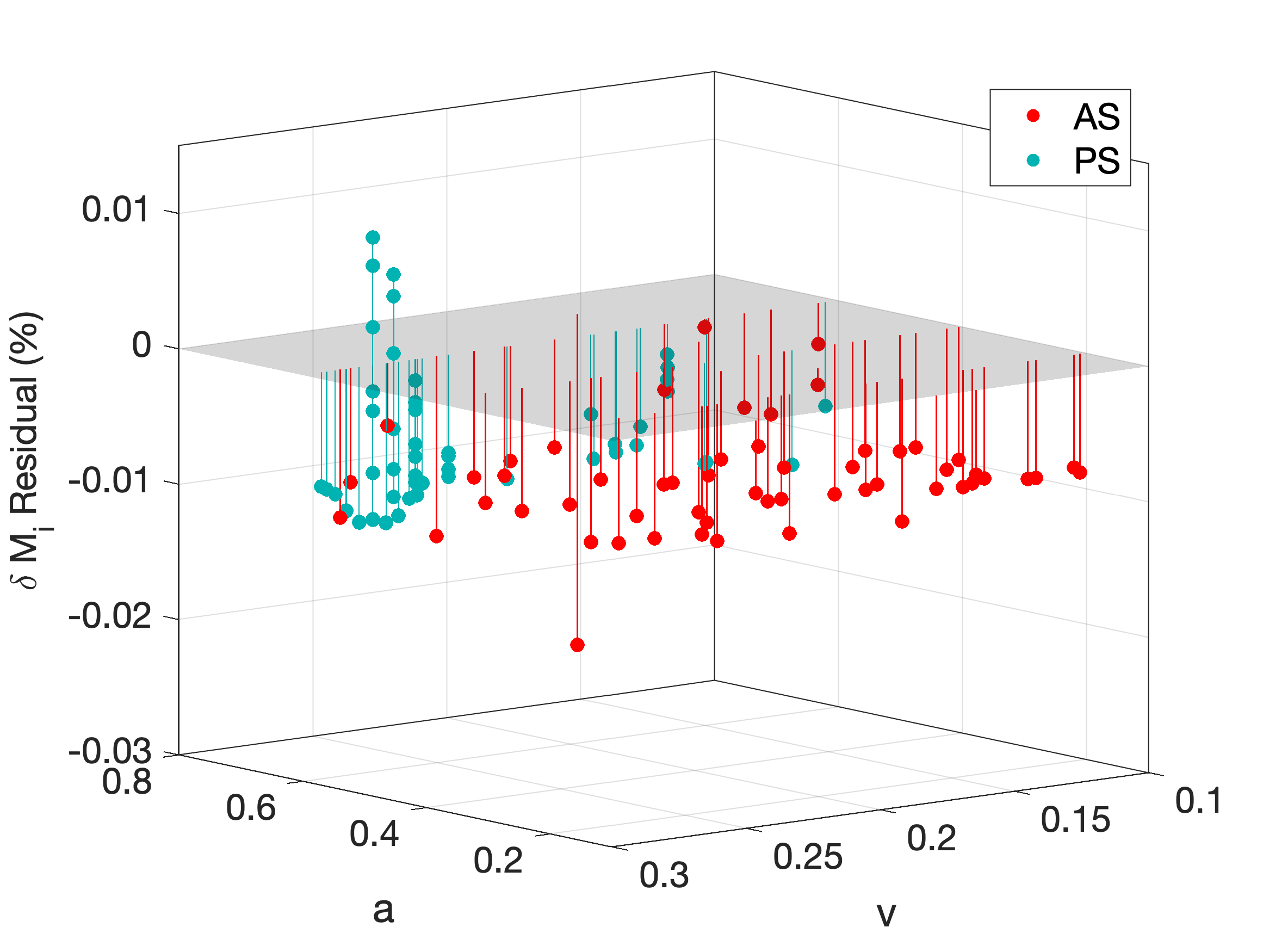} }
 \subfloat[]{     \includegraphics[width=0.47\textwidth]{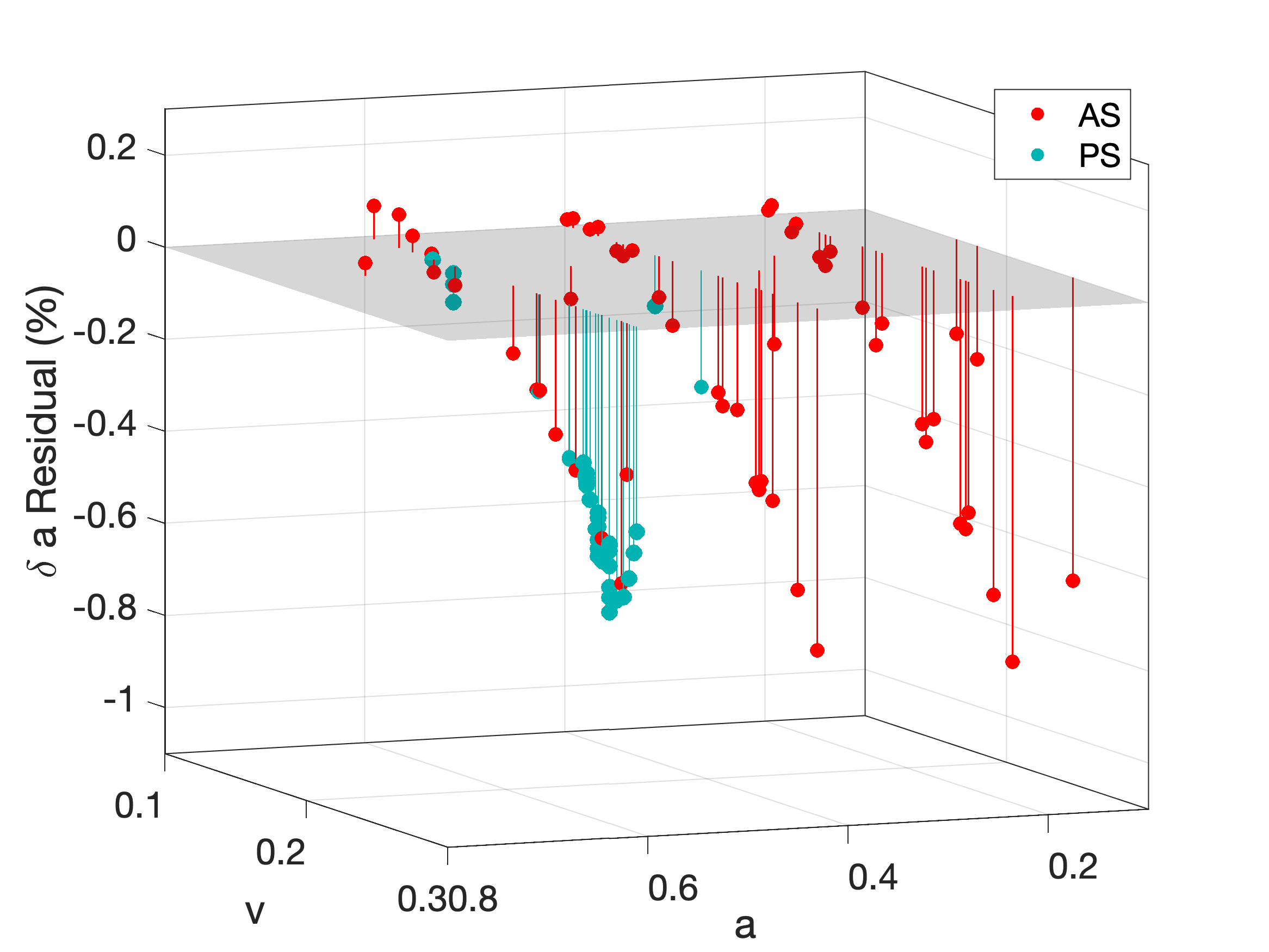}}

    \caption{A comparison between the percent changes (a) $\delta M_i$ and (b) $\delta a$ in \bh{} irreducible masses in \bbh{} simulations (red  for aligned spins and green  for precessing binaries) with the single \bh{} fit (grey surface). Residuals $:=$ (data - fit) between the single \bh{} fit and the \bbh{} data are shown in (c) and (d).}
    \label{fig:fit_mass}
\end{figure*}

\emph{Connecting with binary black holes simulations:}
Once we quantified the changes in mass and spin for the single \bh{} simulations, the next step was to investigate whether these changes are the same as those observed in each of the holes in \bbh{} simulations. We looked at  107 binary simulations in our catalogue: 67 precessing binaries, and 40 aligned spin, non-precessing binaries. Figures~\ref{fig:fit_mass}a and~\ref{fig:fit_mass}b show with points $\delta M_i$ and $\delta a$ for the \bh{s} in the \bbh{s} (red for aligned spins and green points for precessing binaries). Grey surfaces denote the single \bh{} fit. Figures~\ref{fig:fit_mass}c and~\ref{fig:fit_mass}d show the corresponding residuals. The residuals in the masses are $O(10^{-2})\,\%$, and for $v \le 0.24$  the spin residuals are $O(10^{-1})\,\%$. On the other hand, for $v \gtrsim 0.24$ the spin residuals are $O(1)\,\%$ and all negative. Since the residuals are data-fit, this implies that the single \bh{} fit overestimates the effect of junk radiation for $v \gtrsim 0.24$. For reference, \bh{s} with $v \approx 0.24$ involve binary systems with initial separations of $\approx 10\,M$ and gravitational wave frequency $\omega\,M \approx 0.048$.  The levels of residuals for $\delta M_i$ and for $\delta a$ if $v \lesssim 0.24$ give us confidence that the fitting formula derived from single \bh{} simulations provides good estimates applicable to binary simulations. 

\emph{When to worry about junk radiation:} Finally, we present a couple of examples of how the single \bh{} junk radiation fits can be used in \bbh{} simulations. From the waveforms in these simulations, we quantify whether one needs to worry about the effects of junk radiation in gravitational wave analysis for LIGO and LISA sources. 

Assuming that one wants to simulate a binary with \bh{s} having irreducible masses $M_i^{1,2}$, spins $a^{1,2}$ and speeds $v^{1,2}$, the task is to find irreducible masses $\bar M_i^{1,2}$ and spins $\bar a^{1,2}$ to use in the initial data such that the junk radiation modifies these values and yields the desired values $M_i^{1,2}$ and $a^{1,2}$. The adjusted values  $\bar M_i^{1,2}$ and  $\bar a^{1,2}$ can be found by solving the following equations:
\begin{eqnarray}
M_i^{1,2} &=& [1+\delta \bar M_i^{1,2}(\bar a^{1,2}, v^{1,2})]\,\bar M_i^{1,2}\label{eq:d_mass}\\
a^{1,2} &=& [1+\delta \bar a^{1,2}(\bar a^{1,2}, v^{1,2})]\,\bar a^{1,2}\label{eq:d_spin}
\end{eqnarray} 
where the junk radiation changes in the formulas above are given as fractional changes not percentages. As mentioned before, for \bh{s} moving with speeds $\gtrsim 0.24$, the formulas overestimate the correction on the spin magnitude. For those cases, we estimate that only a 10\% correction should be applied. 

\begin{figure}
    \centering
    \includegraphics[width=0.45\textwidth]{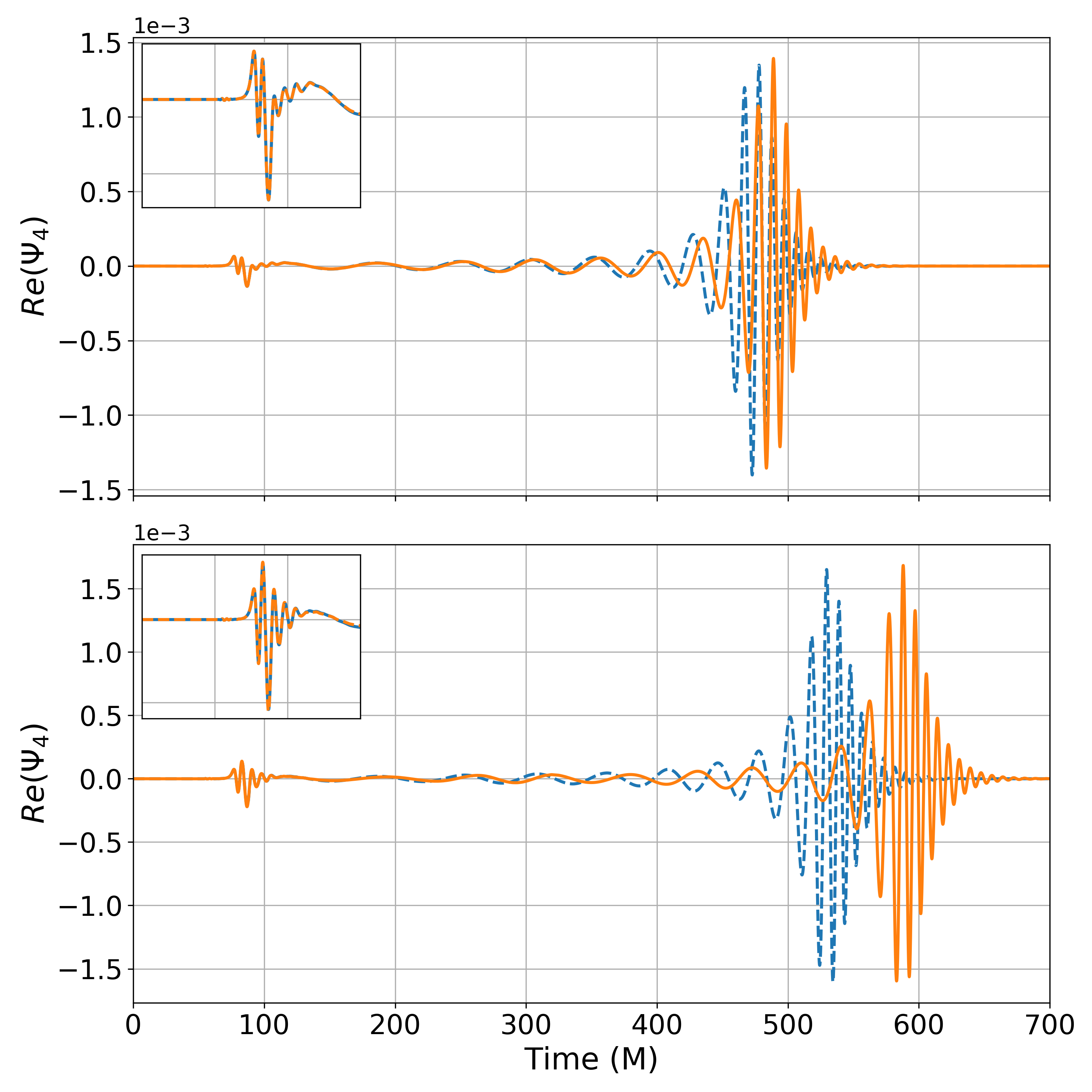}
    \caption{Mode (2,2) of $\Psi_4$ for two equal mass, aligned spin binaries. Top panel shows the cases with $a\approx 0.6$ and the bottom for $a\approx 0.8$. In blue is the waveform from the \emph{raw} simulation and in orange the waveform in which the masses and spins have been adjusted according to the single \bh{} fitting formulas. Insets focus on the junk radiation.}
    \label{fig:waveforms}
\end{figure}

\begin{table}[]
\begin{tabular}{|c|c|c|c|c|c|c|}
\hline
Type 
&\multicolumn{3}{c|}{$t=0$}
&\multicolumn{3}{c|}{$t>t_\text{junk}$}\\
\hline
& $M_h/M$ &  $M_i/M$ & $a$ &$M_h/M$ &$M_i/M$ & $a$  \\
\hline\hline
 raw         &0.4994  & 0.4736  &  0.6013  &0.5004 & 0.4744 &  0.6032  \\
 adj            & 0.4985    & 0.4729  &  0.6000  & 0.4994&0.4736 &  0.6018  \\
 raw            &  0.4972  & 0.4430  &  0.8090  & 0.5001& 0.4471 &  0.8013  \\
 adj             &   0.4939 & 0.4387  & 0.8161  & 0.4971& 0.4431  &  0.8079 \\
\hline
\end{tabular}
 \caption{Mass of the \bh{,} its irreducible mass and spin magnitude at the beginning of the simulation and after the junk radiation has dissipated.}
\label{table:binaries}
\end{table}

Figure~\ref{fig:waveforms} depicts the $l=2$, $m=2$ mode of Weyl scalar $\Psi_4$ for two pairs of simulations. All four simulations consist of equal mass, aligned spin binaries. The top panel shows the cases with $a\approx 0.6$ and the bottom for $a\approx 0.8$. In blue is the waveform from the \emph{raw} simulation and in orange the waveform in which the masses and spins have been adjusted according to the single \bh{} fitting formulas. The mass of the \bh{,} its irreducible mass and spin magnitude at the beginning of the simulation and after the junk radiation has dissipated are given in Table~\ref{table:binaries}. Notice that with the junk adjustment $M_i/M (t > t_\text{junk};\text{adj}) \approx M_i/M (t=0;\text{raw})$
and $a(t > t_\text{junk};\text{adj}) \approx a(t=0;\text{raw})$, as needed.

\begin{figure}
    \centering
  \includegraphics[trim = {20 220 0 250}, width=0.45\textwidth]{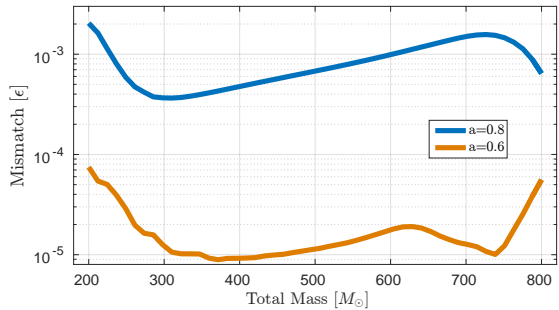}
    \caption{Mismatch as a function of total-mass of the binary for waveforms from the \emph{raw} and adjusted with masses and spins. The blue cover refers $a\approx 0.6$ and the yellow is for $a\approx 0.8$. The mismatch has been computed for Advanced LIGO design noise and masses are kept in the detector frame.}
    \label{fig:mismatch}
\end{figure}

The mismatches $\epsilon$ between the waveform from \emph{raw} and adjusted simulations in Advanced LIGO are shown in Figure~\ref{fig:mismatch}. For  $a\approx 0.8$, $\epsilon \sim10^{-3}$ in average, while for $a\approx 0.6$, $\epsilon \sim10^{-5}$. To avoid astrophysical inference with a bias, one needs $\epsilon \,\rho^2 \lesssim  1$~\cite{2008PhRvD..78l4020L}. This implies that waveforms of sources with highly spinning \bh{} measured with $S/N$ of $\rho \gtrsim 30$ will exhibit inference biases if they are not corrected for the junk radiation. For future detectors such as the Einstein Telescope and LISA, even low spin sources with $\rho \gtrsim 10^2$ would require junk radiation correction.

\emph{Conclusion:} Using simulations of single punctures with different spins and linear momentum, we have investigated the effect of junk radiation on the mass and spin of the \bh{.} We found that junk radiation does not affect the direction of the spin. With these results, we obtained fitting functions that can be used to predict changes in the masses and spin magnitudes of \bh{s} in binary systems. We tested the effectiveness of the  fitting functions and showed that for binary systems with highly spinning \bh{s}, inference biases would be introduced by waveforms that not are corrected for junk radiation in sources with SNRs $ \gtrsim 30$.

\emph{Acknowlegements:} Work supported by  NSF grants 1806580, 1809572, 1550461, and XSEDE  allocation TG-PHY120016.

\bibliography{ms}

\end{document}